\documentstyle[epsf]{aipproc}

\begin{document}

%remove for camera-ready copy
\begin{flushright}UMN-D-00-4 \\ July 2000 \end{flushright}

\title{Methods for the Nonperturbative Approximation
of Form Factors and Scattering Amplitudes%
%remove footnote for camera-ready copy
\footnote{To appear in the proceedings of 
the 7th Conference on the Intersections of
Particle and Nuclear Physics, Qu\'ebec City, May 22-28, 2000.}
}

\author{John R. Hiller%
%remove footnote for camera-ready copy
\footnote{\baselineskip=14pt
Work supported in part by the Department of Energy,
contract DE-FG02-98ER41087.}%
}

\address{Department of Physics \\
University of Minnesota Duluth \\
Duluth Minnesota 55812}

\maketitle

\begin{abstract}
Methods are described for the nonperturbative calculation of wave 
functions and scattering amplitudes in light-cone quantization.  
Form factors are computed from the boost-invariant wave functions, 
which appear as coefficients in a Fock-state expansion of the 
field-theoretic eigenstate.  A technique is proposed for calculating 
scattering amplitudes from matrix elements of a $T$ operator between 
such composite-particle eigenstates.
\end{abstract}

\section*{Introduction}

To benefit from the recent progress on the calculation of
field-theoretic bound states in light-cone 
quantization \cite{jhiller:review,jhiller:PV}, we explore 
methods by which form factors and scattering amplitudes can be 
extracted nonperturbatively.  In the case of form factors, this is
relatively straightforward; well-known formulas \cite{jhiller:Drell}
yield the form factors as overlap integrals of Fock-state
wave functions.  For scattering amplitudes, the way is less
certain.  One possible method \cite{jhiller:QCD00} is discussed
briefly here.  Others have been considered by Kr\"oger
\cite{jhiller:Kroger}, Ji and Surya \cite{jhiller:JiSurya}, and 
Fuda \cite{jhiller:Fuda}.

The formulations given are in terms of light-cone coordinates
\cite{jhiller:Dirac,jhiller:review}, where $x^+\equiv t+z$ plays 
the role of time and the conjugate variable $p^-\equiv E-p_z$ is 
the light-cone energy.  The light-cone three-momentum is
$\underline{p}=(p^+\equiv E+p_z,{\bf p}_\perp)$.  An
eigenstate $|P,\sigma\rangle$ of the light-cone Hamiltonian
operators ${\cal P}^\pm$, ${\bf \cal P}_\perp$ and
helicity $\sigma$ is written as a Fock-state expansion
\begin{equation}
|P,\sigma\rangle=\sum_n\int [dx][d^2k_\perp]
              \psi_{P,\sigma}^{(n)}(x,{\bf k}_\perp)
                        |n:\underline{p}_i \rangle\,,
\nonumber
\end{equation}
with
\begin{equation}
\int [dx][d^2k_\perp]=\int\delta(1-\sum_i x_i)
                       \prod_i \frac{dx_i}{\sqrt{x_i}}
           16\pi^3\delta(\sum_i {\bf k}_{\perp i})
                \prod_i \frac{d^2k_{\perp i}}{16\pi^3}
\end{equation}
and where the $\psi^{(n)}$ are wave functions for $n$ particles,
$x_i\equiv p_i^+/P^+$ are longitudinal momentum fractions,
and ${\bf k}_{\perp i}={\bf p}_{\perp i}-x_i{\bf P}_\perp$
are relative transverse momenta.  Use of light-cone coordinates
brings several advantages, including boost invariance of the
wave functions.

The eigenvalue problem ${\cal P}|P,\sigma\rangle=P|P,\sigma\rangle$
for fixed $\sigma$ determines the wave functions as solutions of
a coupled set of integral equations.  A method frequently applied
to these equations is discrete light-cone quantization (DLCQ)
\cite{jhiller:PauliBrodsky,jhiller:review}, which approximates the
integrals by the trapezoidal rule and computes the wave functions 
on an equally spaced momentum grid.  Any bound-state property can
then, in principle, be calculated from these wave functions.
The grid is parameterized by a longitudinal resolution $K$
and transverse resolution $N_\perp$, such that longitudinal
momentum fractions are multiples of $1/K$ and transverse
momenta have as many as $2N_\perp+1$ values in each direction.
The value of $N_\perp$ is associated with a cutoff $\Lambda^2$
on the invariant mass of each constituent and 
with the choice of transverse momentum scale $\pi/L_\perp$.

\section*{Form factors}

For a spin-$1/2$ fermion, the two form factors can be obtained
from matrix elements of the plus component of the electromagnetic
current $J$
\begin{eqnarray}
F_1(Q^2)&=&\frac{1}{2}\langle P+Q,\sigma | J^+(0)/P^+ |P,\sigma\rangle\,, \\
-\left(\frac{Q_x-iQ_y}{2M}\right) F_2(Q^2)&=&
        \frac{1}{4\sigma}\langle P+Q,\sigma | J^+(0)/P^+ |P,-\sigma\rangle\,.
\end{eqnarray}
These can be reduced to overlap integrals \cite{jhiller:Drell}
\begin{equation}
F_1(Q^2) = \sum_n\sum_je_j\int [dx][d^2k_\perp] 
   \psi_{P+Q,1/2}^{(n)*}(x,{\bf k}_\perp^\prime)
   \psi_{P,1/2}^{(n)}(x,{\bf k}_\perp)\,,
\end{equation}
\begin{equation}
-\left(\frac{Q_x-iQ_y}{2M}\right) F_2(Q^2)=
  \sum_n\sum_je_j\int [dx][d^2k_\perp] 
   \psi_{P+Q,1/2}^{(n)*}(x,{\bf k}_\perp^\prime)
   \psi_{P,-1/2}^{(n)}(x,{\bf k}_\perp)\,, 
\nonumber
\end{equation}
in the frame where the photon momentum $Q$ is
written $(0,2Q\cdot P/P^+,{\bf Q}_\perp)$ and
\begin{equation}
{\bf k}_{\perp i}^\prime=\left\{\begin{array}{ll}
       {\bf k}_{\perp i}-x_i{\bf Q}_\perp\,, & i\neq j \\
       {\bf k}_{\perp j}+(1-x_j){\bf Q}_\perp\,, & i=j\,. 
                                        \end{array}\right.
\end{equation}

\begin{figure}[h!]
\centerline{\epsfxsize=\columnwidth \epsfbox{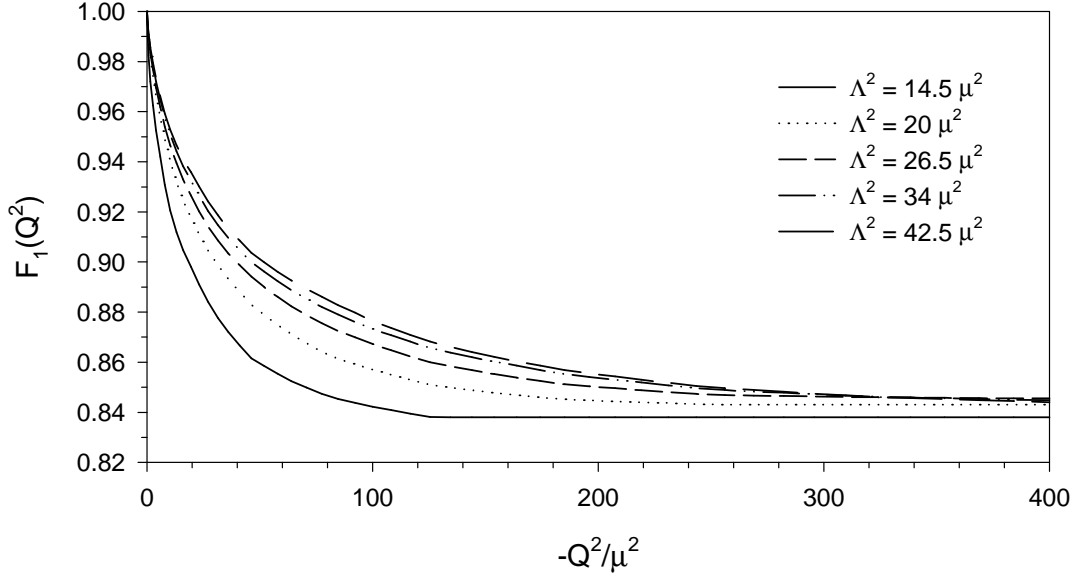} }
%\centerline{\epsfxsize=\columnwidth 
%            \epsfbox{hiller-parallel6.t1.2-formfact.eps} }
\caption{\label{jhiller:fig:FormFactor}
The form factor $F_1$ for fixed longitudinal resolution $K=9$ 
and transverse scale $L_\perp=2\pi/\mu$, and for a particular 
set of model parameters.  Various cutoffs $\Lambda^2$ are 
considered, with the transverse resolution $N_\perp$ ranging 
from 5 to 9.}
\end{figure}

For the model studied by Brodsky, Hiller, and McCartor \cite{jhiller:PV},
an explicit calculation of $F_1$ has been done \cite{jhiller:FormFactor}.
In this model, a bare fermion acts as a source and sink for
bosons of mass $\mu$.  The lowest massive eigenstate is a fermion 
dressed by a boson cloud.  The theory is regulated by a Pauli--Villars
boson \cite{jhiller:PauliVillars} with an imaginary coupling, and
renormalized by fits of physical quantities to ``data.''
Because no spin-flip interactions are included, $F_2$ is
zero.  Results for $F_1$ are shown in Fig.~\ref{jhiller:fig:FormFactor}\@.
The large-momentum-transfer value of $F_1$ is the bare fermion
probability and therefore is not zero.  

\section*{Scattering amplitudes}

The center-of-mass cross section for two-body scattering
($A+B\rightarrow C+D$) is \cite{jhiller:Peskin}
\begin{equation}
\frac{d\sigma}{d\Omega}_{\rm cm} =\frac{1}{2E_A 2E_B v_{\rm rel}}
              \frac{|\vec{p}_C||{\cal M}_{fi}|^2}{16\pi^2 E_{\rm cm}}\,,
\end{equation}
where ${\cal M}_{fi}$ is the invariant amplitude obtained from
the $S$ matrix
\begin{equation}
S_{fi}=\langle f|i\rangle+(2\pi)^4 \delta^{(4)}(p_f-p_i)i{\cal M}_{fi}
      =\delta_{CD,AB}-2\pi i\delta(s_{AB}-s_{CD})T_{{\rm LC}fi}\,,
\end{equation}
with $s_{AB}=\frac{m_A^2+p_{A\perp}^2}{p_A^+/P^+}+
                 \frac{m_B^2+p_{B\perp}^2}{p_B^+/P^+}$. 
The $T$ matrix for scattering of composites is given by
\cite{jhiller:QCD00,jhiller:Wick}
\begin{equation}
T_{{\rm LC}fi}=P^+T_{fi}
  =\langle C|V_D^\dagger\frac{1}{s_{AB}+i\epsilon-H_{\rm LC}}V_B|A\rangle
      +\langle C|DV_B|A\rangle\,.
\end{equation}
Here $|A\rangle$ and $|C\rangle$ are composite-particle eigenstates
of the light-cone Hamiltonian $H_{\rm LC}$, and the operator $V_B$ 
is defined by
\begin{equation}
V_B= [H_{\rm LC},B^\dagger]
          -\frac{m_B^2+p_{B\perp}^2}{p_B^+/P^+}B^\dagger\,,
\end{equation}
with $B^\dagger$ the creation operator for the $B$ particle,
{\em i.e.} $|B\rangle=B^\dagger|0\rangle$.  This construction
generalizes one presented some time ago by Wick \cite{jhiller:Wick}.
Details can be found in Ref.~\cite{jhiller:QCD00}.  Given numerical
solutions for the composite-particle eigenstates,
obtained with DLCQ, the most difficult remaining task is the estimation
of the matrix element of $(s+i\epsilon-H_{\rm LC})^{-1}$.
For this type of matrix element, the recursion method
of Haydock \cite{jhiller:Haydock} has worked well.  A nonrelativistic 
application is described in \cite{jhiller:QCD00}; an application
to the field-theoretic model studied in \cite{jhiller:PV} is in progress.

\section*{Acknowledgments}
This work was supported in part by the Minnesota Supercomputing Institute
through grants of computing time and by the Department of Energy,
contract DE-FG02-98ER41087.

\end{document}